# Extent of Variation Resilience in Strained CMOS: From Transistors to Digital Circuits

[1]Ahmad Ehteshamul Islam, Charles Augustine, Kaushik Roy, and Muhammad Ashraful Alam

School of ECE, Purdue University, West Lafayette, IN, USA

## Abstract

Process-related and stress-induced changes in threshold voltage are major variability concerns in ultra-scaled CMOS transistors. The device designers consider this variability as an irreducible part of the design problem and use different circuit level optimization schemes to handle these variations. In this paper, we demonstrate how an increase in the negative steepness of the universal mobility relationship improves both the process-related (*e.g.*, oxide thickness fluctuation, gate work-function fluctuation), as well as stress-induced or reliability-related (*e.g.*, Bias Temperature Instability or BTI) parametric variation in CMOS technology. Therefore, we calibrate the universal mobility parameters to reflect the measured variation of negative steepness in uniaxially strained CMOS transistor. This allows us to study the extent of (process-related and stress-induced parametric) variation resilience in uniaxial strain technology by increasing the negative steepness of the mobility characteristics. Thus, we show that variability analysis in strained CMOS technology must consider the presence of self-compensation between mobility variation and threshold voltage variation, which leads to considerable amount of variation resilience. Finally, we use detailed circuit simulation to stress the importance of accurate mobility variation modeling in SPICE analysis and explain why the variability concerns in strained technology may be less severe than those in unstrained technology.

---

[1] Email: aeislam@ieee.org, Phone: 765-494-5988, Fax: 765-494-2706

## I.  INTRODUCTION

The scaling of CMOS technology based on Moore's law has so far concentrated on improving the performance of transistor by reducing dimensions (*e.g.*, effective oxide thickness *EOT*, channel length *L*), as well as by incorporating new materials within the transistor structure (*e.g.*, metal for gate, high-κ materials for dielectric and strained-silicon for channel). Such aggressive scaling has the unintended consequence of making the statistical fluctuation of transistor parameters increasingly important [1]. Specifically, current CMOS transistors suffer from statistical variation in process parameters [2-4] like *EOT*, *L*, substrate dopant ($N_D$), gate-substrate work-function difference ($\Phi_{GS}$), *etc.* These effects give rise to time-zero variation in performance parameters (like threshold voltage $V_T$, drain current $I_D$, *etc.*) among transistors of a particular technology. Moreover, use of high oxide electric field ($E_{ox}$) and high-κ dielectric (like SiON, HfO$_2$, *etc.*) have introduced additional concerns from time-dependent variation in transistor parameters due to defect formation within the bulk of the oxide or at the substrate/dielectric interface. These lead to significant reliability concerns in CMOS technology (*e.g.*, Positive Bias Temperature Instability or PBTI for NMOS [5], Negative Bias Temperature Instability or NBTI for PMOS [6]).

Until now, the research on transistor's parametric variations have focused on correlating the fluctuation in parameters to variation in device characteristics [3, 7, 8], as well as mitigating these variations through the use of extra processing steps (*e.g.*, use of monolayer doping for reducing dopant fluctuation [9], plasma nitridation over thermal nitridation in SiON CMOS technology for reducing time-dependent $V_T$ variation [10]). In spite of these efforts, device designers often feel that process modifications alone can not adequately address the variation problem without unacceptable loss in device performance. So a standard option is to operate the transistors with an extra guard band voltage, over and above the voltage that is required for nominal operation [11, 12], so that the circuit remains functional despite significant parametric variation. Similarly, there are proposals for using adaptive body bias [13-15], adaptive power supply (*i.e.*, circuit sleeping) [16], (area-)resizing of transistors along the critical path [17], *etc.* for minimizing the impact of temporal variability or reliability. Such circuit level reliability optimization requires one to monitor the degradation level of IC at different stages of operation using specialized circuits like 'Silicon Odometers' [11] (based on the measurement of quiescent leakage current [18], for example).

Recently, we have discussed the possibility of self-compensation of NBTI in strained PMOS transistors [19] and PBTI in high-κ NMOS transistor [20, 21]. Sufficient negative steepness in the effective mobility ($\mu_{eff}$) of the inversion carrier vs. vertical effective electric field ($E_{eff}$) characteristic, achieved through uniaxial strain in [19], has been the key requirement for designing such *self-compensated or variation-resilient transistors*. A steep $\mu_{eff}$-$E_{eff}$ characteristic ensures that NBTI and PBTI in these transistors result in an opposing fluctuation for $\mu_{eff}$ and carrier density ($N_{inv}$) within the inverted PMOS channel. As a result, the drivability (*i.e.,* drain current $I_D \sim \mu_{eff}N_{inv}$) of a variation-resilient transistor becomes less sensitive to NBTI and PBTI (hence, BTI as a whole). This beneficial effect has also been corroborated in the simple circuit analysis reported in [20, 21].

In this paper, we broaden the discussion of self-compensation in *four important ways*. **First**, we show that the principles of self-compensation is relevant to different process variations, as well and show that self-compensation is indeed possible for variations related to different oxide and gate parameters (*e.g.*, *EOT*, $\Phi_{MS}$); as such, the 'degradation-compensated' strained transistor proposed in [19] can also be considered as a 'variation-resilient' transistor. **Second**, we characterize the $\mu_{eff}$-$E_{eff}$ relationship as a function of uniaxial strain (presumably for the first time) and identify the *practical* limits of uniaxial strain technology in achieving self-compensation. **Third**, we highlight the *limitations* of achieving self-compensation in modern short channel transistors by showing its inability to reduce the effects of channel length variation, random dopant fluctuation (RDF), as well as variation in off-state leakage current. And **finally**, we verify the consequence of self-compensation in a set of digital circuits and show that the delay degradation in digital circuits can indeed be reduced by considering physical changes in transistor parameters (like $V_T$, $E_{eff}$, $\mu_{eff}$, *etc.*) during circuit simulation.[2] Therefore, we illustrate how variation resilience may ease the burden of designers in guard-banding against some of the variability issues (like BTI, *EOT*-fluctuation, $\Phi_{MS}$-fluctuation) that the nanoscale CMOS technology is currently struggling to handle.

---

[2] Note that conventional SPICE simulation will fail to capture the effect of self-compensation, due to the reasons discussed in section VII-A.

## II. PRINCIPLE OF VARIATION RESILIENCE

Let us first review the design principle involving self-compensation in a variation-resilient transistor and compare its characteristics with a classical CMOS transistor. For this, we need to understand the effect of process-related and stress-induced parametric variation in transistors drain current. Drain current in linear region ($I_{D,lin}$) can be expressed as $I_{D,lin} \sim \mu_{eff} N_{inv}$; where, $N_{inv} \sim (V_G - V_T)$. Thus, the fluctuation of $I_{D,lin}$ ($\Delta I_{D,lin}$) can be expressed as:

$$\Delta I_{D,lin} / I_{D,lin} \sim \Delta \mu_{eff} / \mu_{eff} + \Delta N_{inv} / N_{inv} \sim \Delta \mu_{eff} / \mu_{eff} - \Delta V_T / (V_G - V_T). \qquad (1)$$

According to (1), a transistor can be resilient to $I_{D,lin}$ variation (*i.e.*, have $\Delta I_{D,lin} \sim 0$) *only if* an increase (decrease) in $\mu_{eff}$ compensates the corresponding increase (decrease) in $V_T$. All MOS transistors have negative differential steepness in its $\mu_{eff} - \underline{E_{eff}}$ characteristics (Fig. 1b), mainly due to effect of surface roughness [22] and this 'negative steepness' dictates that any increase in $V_T$ is always accompanied by an increase in $\mu_{eff} @ V_G$ and vice versa. [3]

In classical MOS transistor, the steepness of the $\mu_{eff} - E_{eff}$ or $\mu_{eff} - V_G$ (see (10) for $E_{eff} - V_G$ relationship) characteristic is generally insufficient to fully compensate the linear increase in $N_{inv}$ with $V_G$. As a result, transfer (*e.g.*, $I_{D,lin}$-$V_G$) characteristic of a classical transistor is dominated by the increase in $N_{inv}$ with $V_G$; as such, a classical transistor have finite ON-state transconductance $g_{m,ON} \equiv \partial I_D / \partial V_G \big|_{V_G \gg V_T}$ (Fig. 1a). In addition, relatively shallow $\mu_{eff} - E_{eff}$ steepness in classical transistors ensures negligible change in $\mu_{eff}$ ($\Delta \mu_{eff}$) due to $\Delta V_T$. Thus according to (1), $\Delta I_{D,lin}$ is dominated by the $\Delta V_T$ term, which makes $I_{D,lin}$ of a classical transistor sensitive to various sources of $\Delta V_T$. To mitigate the effect of $\Delta V_T$ in these transistors, one can use a fraction of the supply voltage ($V_{DD} - V_I$ in Fig. 1a) to guard-band against $\Delta V_T$ [11, 12] − so that even the slowest transistor (at the end of product lifetime) have sufficient drive current ($I_{D,min}$) for a functional integrated circuit.

It is possible however to increase the negative steepness of $\mu_{eff} - E_{eff}$ relationship by various techniques, *e.g.*, by introducing uniaxial strain (thus reducing phonon-scattering, while keeping surface roughness essentially unchanged) [19], by reducing operating temperature (thus making temperature-independent surface roughness scattering dominant over the temperature-dependent phonon-scattering) [21, 23]. And with suitably high negative steepness for the $\mu_{eff} - E_{eff}$ characteristic, positive $\Delta\mu_{eff}@V_G$ due to positive $\Delta V_T$ (and vice versa) can be large enough to balance the two terms in (1), thereby making the variation-resilient transistor insensitive to $\Delta V_T$ (in terms of $I_{D,lin}$). Moreover, since an increase in the negative steepness of $\mu_{eff} - E_{eff}$ or $\mu_{eff} - V_G$ can match the positive steepness of $N_{inv} - V_G$, therefore, variation-resilient transistors have flatter $I_{D,lin}$-$V_G$ characteristics (smaller $g_{m,ON}$) compared to the classical ones (Fig. 1c). Thus consideration of self-compensation between $\mu_{eff}$ and $N_{inv}$ is eventually reflected in flat transfer characteristics above threshold. And it is intuitively clear from Fig. 1c that with such flat transfer characteristics, $I_{D,lin}$ will not be affected by $\Delta V_T$. Therefore, the presence of self-compensation suggests the possibility of reducing the guard-band voltage for $I_{D,min}$.

---

[3] However, in case of interface trap ($N_{IT}$)-related $\Delta V_T$, one needs to consider the decrease in $\mu_{eff} @ E_{eff}$ due to additional coulomb scattering introduced by $N_{IT}$ [22]; as a result, overall change in $\mu_{eff} @ V_G$ can be either positive or negative depending on the $\mu_{eff} - E_{eff}$ steepness [19].

Now, let us see whether the presence of $I_{D,lin}$ self-compensation ensures corresponding behavior in the saturation regime, as well. Here, we theoretically analyze the effect of self-compensation in saturation region drain current ($I_{D,sat}$) in the following way. According to scattering theory [24], modern transistors operate in quasi-ballistic regime and $\Delta I_D$ in linear and saturation regions can be approximated as [21, 24],

$$\left.\frac{\Delta \boldsymbol{I}_{D,lin(sat)}}{\boldsymbol{I}_{D,lin(sat)0}}\right|_{V_G} = \left(1 - \boldsymbol{B}_{lin(sat)}\right)\left.\frac{\Delta\mu_{eff}}{\mu_{eff\,0}}\right|_{V_G} - \frac{\left|\Delta\boldsymbol{V}_{T,lin(sat)}\right|}{\boldsymbol{V}_G - \boldsymbol{V}_{T,lin(sat)0}}. \quad (2)$$

Now, the ballistic coefficients in the linear regime ($B_{lin}$) is ~ 0, in the saturation regime ($B_{sat}$) ranges from $0 << B_{sat} < 1$ [24, 25]. Therefore, one expects self-compensation to be more effective for $\Delta I_{D,lin}$ compared to $\Delta I_{D,sat}$.

Finally, let us consider the implication of self-compensation on the variation of $I_{OFF}$ in the sub-threshold region ($V_G < V_T$). Since $I_{OFF}$ for a transistor is dominated by thermionic emission (when band-to-band tunneling is negligible[4]), we can write $I_{OFF} \sim exp \ [-q|V_T|/mk_BT]$ [27] and its differentiation in the sub-threshold regime results –

$$\frac{\Delta I_{OFF}}{I_{OFF}} = \frac{1}{mk_BT}\left(-\left|\Delta V_T\right| + \frac{\Delta m}{m}\left|V_T\right|\right) \tag{3}$$

where, $m = 1 + \left(C_D + C_{IT}\right)/C_{di}$ is the body-effect co-efficient, $C_D$ is the depletion-layer capacitance, $C_{IT}$ is the interface trap capacitance, and $C_{di} = \varepsilon_{SiO_2}/EOT = \varepsilon_{di}/T_{PHY}$ is the dielectric capacitance, $\varepsilon_{SiO_2}$ is the relative permittivity of SiO$_2$, and $\varepsilon_{di}$ , $T_{PHY}$ are relative permittivity and thickness of high-κ gate dielectric. Since, $\Delta m \sim \Delta V_T/E_{g,Si} \sim \Delta V_T$ (where, $E_{g,Si} \sim 1eV$ is the silicon bandgap), (3) suggests that a partial compensation between $\Delta V_T$ and $\Delta m$ (which obviously has different physics compared to the self-compensation between $\Delta\mu_{eff}$ and $\Delta V_T$, discussed so far) will only be possible for $|V_T|/m \sim 1$. However, modern transistors has $m > 1$ and $|V_T| \sim 0.3V$ [28]. Therefore, self-compensation between $\Delta\mu_{eff}$ and $\Delta V_T$ has no influence on reducing $\Delta I_{OFF}$ in CMOS transistors.

## III. METHOD OF INCREASING μ$_{EFF}$-E$_{EFF}$ STEEPNESS

We have now established larger $\mu_{eff}$-$E_{eff}$ steepness or smaller $g_{m,ON}$ as a requirement for designing self-compensated transistors. Such increase in $\mu_{eff}$-$E_{eff}$ steepness can be achieved by having a larger increase in $\mu_{eff}$ at low $E_{eff}$ compared to the $\mu_{eff}$ at high $E_{eff}$. Within the picture of universal mobility, $\mu_{eff}$ can be expressed as [22]:

$$\mu_{eff}^{-1} = \mu_{coul}^{-1} + \mu_{ph}^{-1} + \mu_{sr}^{-1} \tag{4}$$

where, $\mu_{coul}$, $\mu_{ph}$, and $\mu_{sr}$ are components of $\mu_{eff}$ due to coulomb scattering, phonon scattering and surface roughness scattering, respectively. As shown in [22], $\mu_{coul}$, $\mu_{ph}$ are the dominant component at low $E_{eff}$ and $\mu_{sr}$ is the dominant component at high $E_{eff}$. Thus to achieve self-compensation by increasing $\mu_{eff}$-$E_{eff}$ steepness, one needs to improve $\mu_{coul}$, $\mu_{ph}$ and keep similar $\mu_{sr}$ by increasing uniaxial strain within the channel [19] or by reducing operating temperature [21]. In the following, we qualitatively (section A) and quantitatively (section B) explain the variation in $\mu_{eff}$-$E_{eff}$ steepness as a function of uniaxial strain, thus identify the extent of uniaxial strained technology necessary for self-compensation of $\Delta V_T$ (section C).

---

[4] If $I_{OFF}$ is dominated by band-to-band tunneling (as the case for short-channel transistors having low bandgap Ge substrates [26]), any change in $V_T$ will directly reflect in a corresponding change in $I_{OFF}$. Hence, self-compensation for $I_{OFF}$ is also unexpected in such short-channel transistors.

## A.    Qualitative Theory

As widely used for transistor's performance improvement [43], uniaxial strain (rather than biaxial strain) is also preferable for achieving self-compensation. Biaxial tensile strain though shows performance or $\mu_{eff}$ improvement for NMOS, compared to that for PMOS [43], it will not be suitable for obtaining self-compensation in NMOS transistors. This is because biaxial tensile strain increases $\mu_{sr}$ (as well as $\mu_{ph}$) for NMOS by reducing the roughness amplitude [29, 30]. Thus biaxial tensile strain in NMOS improves $\mu_{eff}$ at all $E_{eff}$, and hence does not increase the $\mu_{eff}$-$E_{eff}$ steepness [30]. On the other hand, process induced uniaxial strain by source/drain engineering is introduced after the gate insulator growth. Thus, it does not change the Si and gate insulator interface properties and hence results negligible change in surface roughness or $\mu_{sr}$ [31]; thereby, uniaxial strain satisfies the first requirement (*i.e.*, reduced increase in $\mu_{eff}$ at high $E_{eff}$) of self-compensation.

Now consider if the requirements of enhanced $\mu_{coul}$ and $\mu_{ph}$ (for self-compensation) are also satisfied in uniaxial strain technology. We consider the consequence of using tensile strain and <110> compressive strain for NMOS and PMOS, respectively; because these uniaxial strains improve transistor performance significantly [43]. *Tensile strain* splits the $\Delta_2$-$\Delta_4$ electron valleys and hence reduces inter-valley phonon scattering or increases $\mu_{ph}$ for electron transport. In addition, since $\Delta_2$ constitutes the lower electron valley under tensile strain, there is also an increase in the occupancy of electron at the $\Delta_2$ electron valley having lower transport effective mass. As such, tensile strain also increases $\mu_{coul}$ for electron transport [32]. As a result, self-compensation in NMOS transistor can be achieved through tensile strain. On the other hand, *compressive strain* splits the heavy hole-light hole (HH-LH) valleys and hence reduces inter-valley phonon scattering or increases $\mu_{ph}$ for hole transport. In addition, since HH valley has the higher occupancy of holes under compressive strain (which also has lower transport effective mass in the <110> direction), <110> compressive strain also expected to increase $\mu_{coul}$ for hole transport. As a result, self-compensation in PMOS transistor can be achieved through compressive strain.

Here, we experimentally demonstrate the consequence of <110> compressive uniaxial strain in achieving self-compensation for PMOS transistor. Since the theory of self-compensation is similar for NMOS and PMOS transistors (see section III-A), we also expect similar improvement in NMOS transistor, as well. The PMOS transistors under study have <110> compressive uniaxial strain applied through SiGe Source/Drain and contact etch stop layer (therefore, a reduction of $L$ increases strain [33]); the transistors also have a doping density of $N_{Dop}$ ~ $3 \times 10^{17}$ cm$^{-3}$ and $EOT$ ~ 1.4nm. Now, to estimate the effect of uniaxial strain ($\varepsilon$) on $\mu_{eff}$-$E_{eff}$, we determine $\mu_{eff}$ and $E_{eff}$ for unstrained and strained transistors having different channel length ($L$), using the measurement procedure similar to [34]. Then, we calculate the mobility enhancement factor $\mu_{eff,\varepsilon}/\mu_{eff,\varepsilon=0}$ as a function of $E_{eff}$ at different $L$ (or equivalently at different $\varepsilon$).[5] Finally, by using $\mu_{eff,\varepsilon}/\mu_{eff,\varepsilon=0}$ and $\mu_{eff,\varepsilon=0}$ for longest $L$ transistor (thus avoiding the extra mechanism for shorter $L$ transistor [35]), we estimate $\mu_{eff}@E_{eff}$ for different amount of uniaxial compressive strain (Fig. 2). Our measurement of $\mu_{eff}$ *vs.* $E_{eff}$ for unstrained $L = 10\mu m$ transistor is consistent with the universal mobility curve [22]; here, the only difference at low $E_{eff}$ is due to presence of higher substrate doping (~$3 \times 10^{17}$ cm$^{-3}$) in our transistors. Fig. 2 also indicates an increase in $\mu_{eff}@E_{eff}$ with strain, as well as an increase in the negative steepness of the $\mu_{eff}$ *vs.* $E_{eff}$ characteristics, as expected from the qualitative analysis in section A.

To capture the strain dependence in a quantitative mobility model, we add two features in the universal mobility relationship. Based on the discussion in section III-A, we know that compressive uniaxial strain increases $\mu_{coul}$ by having higher occupancy in the lower transport effective mass HH valley and also increases $\mu_{ph}$ due to HH-LH band splitting. To capture this, we use the following expressions for $\mu_{coul}$, $\mu_{ph}$, and $\mu_{sr}$ for fitting the $\mu_{eff}$-$E_{eff}$ characteristics in Fig. 2:

---

[5] Here, we could have estimated $\mu_{eff,\varepsilon}/\mu_{eff,\varepsilon=0}$ by measuring $\mu_{eff}$ on transistors having different $L$ (*i.e.*, $\mu_{eff}$ for longest $L$ transistor will serve as $\mu_{eff,\varepsilon=0}$ and $\mu_{eff}$ for shorter $L$ transistors will serve as $\mu_{eff,\varepsilon}$). However, such procedure requires explicit decomposition between the effect of strain and the effect of extra scattering from source/drain halo regions [35] for shorter channel transistors. Therefore, we estimate $\mu_{eff,\varepsilon}/\mu_{eff,\varepsilon=0}$ at a particular $\varepsilon$ (or $L$), by comparing $\mu_{eff}@E_{eff}$ of unstrained and strained transistors having same channel length.

$$\mu_{coul} = \frac{\mu_1 \mu_{1str}}{\left[\dfrac{N_{Dop}}{10^{18}} + \beta_{IT} \dfrac{N_{IT}}{10^{12}} \dfrac{N_{inv}}{10^{12}}\right]} \left(\frac{N_{inv}}{10^{12}}\right)^{\alpha_1} \tag{5}$$

$$\mu_{ph} = \mu_2 E_{eff}{}^{\alpha_2} \left[1 + B\left(\exp\frac{\Delta E_{LH-HH}}{kT} - 1\right)\right] \tag{6}$$

$$\mu_{sr} = \mu_3 E_{eff}{}^{\alpha_3} \tag{7}$$

Note that the phenomenological expression for strain-induced phonon mobility enhancement for $\mu_{ph}$ in (6) was first introduced in [36] and later used by many groups [37, 38]. Using $\mu_{eff}$-$E_{eff}$ characteristics of unstrained transistor, we first estimate the parameters: $\mu_1 = 97.3$ cm$^2$/V-sec, $\mu_2 = 82.52$ cm$^2$/V-sec, and $\mu_3 = 357.5$ cm$^2$/V-sec. The remaining parameters are consistent with literature; *vis.* $\alpha_1 = 1$, $\alpha_2 = -0.32$ are the same as used in [22] and $\alpha_3 = -1.6$ is the same as extracted from the $\mu_{sr}$ components reported in [39]. Next, band-structure information of strained transistors are used to estimate the effective energy difference of HH-LH valleys in strained ($\Delta_{H,\varepsilon}$) transistors using single sub-band and Airy-function approximation. Similar to the findings in [40], our calculation of $\Delta_{H,\varepsilon}$ shows an increase in $\Delta_{H,\varepsilon}$ with $E_{eff}$ (see Fig. 3a). Then, by comparing $\Delta_{H,\varepsilon}$ for unstrained and strained transistors, we calculate $\Delta E_{LH-HH} = \Delta_{H,\varepsilon} - \Delta_{H,\varepsilon=0}$ at different $E_{eff}$ (see Fig. 3b). Interestingly, $\Delta E_{LH-HH}$ at higher $E_{eff}$ has small deviation from its value of $6.426\varepsilon$ eV at zero $E_{eff}$ (where, 6.426eV is the deformation potential [40]). Using the information of $\Delta E_{LH-HH}$ from Fig. 3b, we fit $\mu_{eff}$-$E_{eff}$ characteristics (Fig. 2) and hence estimate $\mu_{1str} = (1+9\times10^4\varepsilon)$, $B = 0.27$. Finally, for studying the effect of $N_{IT}$ in section VII-B, we use $\beta_{IT} \sim 0.04$ for the wavefunction interaction parameter [41], which has been obtained by fitting $\mu_{eff}$-$E_{eff}$ characteristics before and after $N_{IT}$ generation [34] and is assumed to be strain independent (as observed in [42]).

### C.    Self-Compensation at Practical Strain Limit

Using the strain-dependent mobility model of (4)-(7), we estimate $\mu_{eff}$-$E_{eff}$ for different levels of strain (Fig. 4a), which suggests that for higher strain (*i.e.*, for $\varepsilon > 2.5\%$ in Fig. 4a), mobility enhancement gets limited by the strain independent $\mu_{sr}$ component.[6] Next in Fig. 4b, we calculate $I_{D,lin}$-$V_G$ for $L = 100$nm using $\mu_{eff}$-$E_{eff}$ of Fig. 4a. Fig. 4b suggests that with $\varepsilon = 2.5\%$ flatness or total self-compensation in $I_{D,lin}$-$V_G$ characteristics can be achieved at $V_{GS}\sim 1.2$-$1.5$V. However, for a practical transistor, one needs to consider the contribution from additional scattering mechanisms near source/drain halo regions [35], which may modify this estimate of $\varepsilon$ in achieving self-compensation. Thus we conclude that complete self-compensation ($\Delta I_{D,lin} \sim 0$ at all $V_G$) may not be possible within practical strain limits. However, as discussed in section VII-B, *partial* self-compensation (through positive $\Delta\mu_{eff}$) is indeed a reality in uniaxial strained transistors and circuits.

---

[6] The observation of mobility enhancement saturation through uniaxial strain is also consistent with [43].

## IV. SOURCES OF $V_T$ FLUCTUATION

From our discussions in section III, we know that it is possible to achieve partial self-compensation through uniaxial strain. In this section, using the following expression for $V_T$ in bulk CMOS [27], we identify the sources of $\Delta V_T$ that one needs to self-compensate:

$$V_T = \Phi_{GS} + \frac{EOT}{\varepsilon_{SiO_2}}\sqrt{4k_B T \varepsilon_{Si} N_{Dop}\ln\frac{N_{Dop}}{n_i}} + \frac{2k_B T}{q}\ln\frac{N_{Dop}}{n_i} + q\frac{N_T(t)}{C_{di}}, \quad (8)$$

In (8), $N_T(t)$ indicates the time-dependent contribution from defects. These defects can be present within the oxide and these oxide defects $N_{OT}$ are observed during BTI stress on PMOS/NMOS transistor [5]. Defects can also be present at the oxide/substrate interface and these interface defects $N_{IT}$ are observed during NBTI stress on PMOS transistor [7]. Therefore, (8) indicates that $\Delta V_T$ can arise from fluctuation in process parameters like $\Phi_{GS}$, $EOT$, $N_{Dop}$ or in reliability parameters like $N_T(t)$. As shown in [44, 45], process-related fluctuation in Intel's 45 nm technology corresponds to a $3\sigma$ of $\sim 150$mV, whereas temporal fluctuation has a mean shift of $\sim 60$mV with a $3\sigma$ of $\sim 30$mV. In sum, one needs to handle a $\Delta V_{T,max} \sim 240$mV (above $V_{T0}$) by designing transistors to operate at a voltage higher than $V_1$.

## V. RESILIENCE TO PROCESS FLUCTUATIONS

Given the principle of $I_{D,lin}$ self-compensation discussed in section II, let us explore in this section how an appropriately designed variation-resilient transistor (through uniaxial strain) might lead to $I_{D,lin}$ self-compensation against process fluctuations related to $\Phi_{GS}$ and $EOT$. Our approach involves the calculation of $V_T$ using (8) and $V_G$, $E_{eff}$, $I_{D,lin}$, $\mu_{eff}$ using the following equations [27]:

$$V_G = \Phi_{GS} + \left| \sqrt{2\varepsilon_{Si}k_B T N_{Dop}} \left[ \frac{q\psi_S}{k_B T} + \frac{n_i^2}{N_{Dop}^2} \left( e^{q\psi_S/k_B T} - \frac{q\psi_S}{k_B T} - 1 \right) \right] \right| / C_{di} + \psi_S \qquad (9)$$

$$E_{eff} = \frac{1}{\varepsilon_{Si}} \left[ (1-\eta)\sqrt{4k_B T \varepsilon_{Si} N_{Dop} \ln \frac{N_{Dop}}{n_i}} - \eta C_{di} \frac{2k_B T}{q} \ln \frac{N_{Dop}}{n_i} + \eta \frac{\varepsilon_{SiO_2}}{EOT}(V_G - \Phi_{MS}) \right] \qquad (10)$$

$$I_{D,lin} = \mu_{eff} C_{di} \frac{W}{L} V_{DS} \frac{\dfrac{2mkT}{q} \ln \left[ 1 + \exp \dfrac{q(V_G - V_T)}{2mkT} \right]}{1 + \dfrac{2m}{m-1} \exp \dfrac{q(V_T - V_G)}{2mkT}} \qquad (11)$$

where the symbols have their standard definitions, as in [27]. Note that (11) leads to the usual expressions for $I_{D,lin}$ in sub-threshold (for $V_G << V_T$) and super-threshold (for $V_G >> V_T$) regions. Using (4)-(11), we simulate the transfer characteristics for a PMOS transistor. A unstrained transistor (having $L$ = 130nm, $EOT \sim$ 2.0nm) is prone to ±0.2nm $EOT$ variation (Fig. 5b) and ±10% $\Phi_{GS}$ variation (Fig. 5d). However, increase in strain to 2.5% results in smaller $g_{m,ON}$ near $V_{GS} \sim$ 1.2-1.5V (Fig. 5c, Fig. 5e); therefore, $\Delta V_T$ gives rise to negligible $\Delta I_{D,lin}$ in that voltage range, as discussed earlier.

It is important to understand that while the proposed scheme partially compensates for $EOT$ and $\Phi_{MS}$ fluctuation at certain voltage ranges, unfortunately it can not at all compensate the effects of random-dopant fluctuation (RDF). Indeed, simulations suggest that $I_{D,lin}$ remains sensitive to RDF, irrespective of the $\mu_{eff}$ - $E_{eff}$ steepness (see Fig. 6). To explain this result, consider the expression for $E_{eff}$ in (10). Variation in $N_{Dop}$ mainly effects the first term in the right-hand side of (10). Therefore, any increase (decrease) in $N_{Dop}$, over and above the mean value, raises (lowers) both $V_T$ according to (8), as well as $E_{eff}$ according to (10). Therefore, dopant fluctuation decreases (increases) $\mu_{eff} @ V_G$ with an increase (decrease) in $V_T$ – and as such can not satisfy the requirement for variation resilience based on (1). However, the proposed approach remains relevant as the technology evolves towards fully-depleted silicon-on-insulator with no issues related to RDF. Similarly higher steepness in $\mu_{eff}$ - $E_{eff}$ characteristics can not compensate the effects of channel length fluctuation (LER). This is because changes in $L$ above (below) the mean value decreases (increases) strain and thereby $\mu_{eff}$; as such causes $I_{D,lin}$ to decrease (increase), according to (11), irrespective of $\mu_{eff}$ - $E_{eff}$ steepness.

Thus, we have demonstrated the advantage of increasing $\mu_{eff}$ - $E_{eff}$ steepness (or reducing $g_{m,ON}$) through strain in a variation-resilient transistor for obtaining self-compensation in $I_{D,lin}$ at certain voltage ranges. However, as discussed in (2) and (3), self-compensation is also partially effective for $I_{D,sat}$ and is apparently absent for $I_{OFF}$. However, partial self-compensations in $I_{D,lin}$ and $I_{D,sat}$ also have significant impact in circuit analysis, as discussed in section VII.

## VI. RESILIENCE TO TIME-DEPENDENT FLUCTUATION

Similar to the process-related $V_T$ fluctuations, we can also compensate $N_{OT}$ or $N_{IT}$ -related time-dependent $V_T$ fluctuations by increasing the $\mu_{eff}$ - $E_{eff}$ steepness [19-21]. Based on the same principle, as discussed in section II, $N_{OT}$-induced $\Delta V_T$ will reduce $E_{eff}@V_G$ and hence improve $\mu_{eff}@V_G$, when measured at constant $V_G$. We have experimentally observed such increase in $\mu_{eff}@V_G$ due to $N_{OT}$ in [20, 21], which obviously compensates $\Delta V_T$, according to (1). Similarly, $N_{IT}$-induced $\Delta V_T$ will also reduce $E_{eff}@V_G$ and consequently increase $\mu_{eff}@V_G$ [19]. However, contrary to the cases for $N_{OT}$, $N_{IT}$ introduces additional coulomb scattering centers [22] and hence reduces $\mu_{eff}@E_{eff}$. As a result, $\Delta\mu_{eff}@V_G$ due to $N_{IT}$-induced $\Delta V_T$ is not always positive (as the case for $N_{OT}$-induced $\Delta V_T$) and depends on the $\mu_{eff}$ - $E_{eff}$ steepness [19]. Only for highly strained transistors having larger $\mu_{eff}$ - $E_{eff}$ steepness, $\mu_{eff}$ improvement through $E_{eff}$ reduction dominates over the downward shift of mobility-field curve (*i.e.*, negative $\Delta\mu_{eff}@E_{eff}$) – resulting positive $\Delta\mu_{eff}@V_G$ for $N_{IT}$-induced $\Delta V_T$. As such, according to (1), positive $\Delta\mu_{eff}/\mu_{eff,0}@V_G$ for highly-strained transistors can easily compensate the negative contribution from -$\Delta V_T/(V_G$-$V_{T0})$, resulting negligible $\Delta I_{D,lin}$ for the time-dependent degradation related to $N_{OT}$

## VII. CIRCUIT LEVEL SELF-COMPENSATION

So far we have explored the implication of self-compensation on individual transistor parameters (sections III-V) and demonstrated that partial self-compensation can be routinely achieved under various conditions in modern CMOS transistors involving uniaxial strain. In this section, we study the consequence of self-compensating $N_{IT}$-induced $\Delta V_T$ in digital circuits. First, we show why the current approach of SPICE simulation gives *unphysical* $\Delta E_{eff}$ due to $\Delta V_T$ and hence fails to capture the effect of self-compensation. Later, by considering appropriate changes in $E_{eff}$ due to $\Delta V_T$, we show how self-compensation significantly reduces the circuit-level degradation due to $N_{IT}$, compared to the one obtained from a conventional SPICE analysis (like [46]). Though we restrict our analysis to time-dependent $N_{IT}$ variations, similar analysis should be valid for other sources of $\Delta V_T$ variation.

### A.    Issues with SPICE Analysis: Effect of Unphysical $\Delta E_{eff}$

In current SPICE-based circuit analysis for $N_{IT}$, $\Delta \mu_{eff}$ is calculated either empirically by adding an extra fitting parameter in $\Delta V_T$ estimation [18] or analytically by using the following expression [46] –

$$\mu_{eff} = \frac{\mu_0}{\left(1 + \alpha_{eff} N_{IT}\right) f\left(V_G, V_T\right)} \qquad (12)$$

where $\alpha_{eff}$ is a fitting parameter obtained from mobility degradation experiments performed at constant $E_{eff}$ [47]), and based on the mobility model commonly used in BSIM/SPICE analysis,

$$f\left(V_G, V_T\right) = 1 + \theta_1\left(\frac{V_{gsteff} + 2V_T}{T_{ox}}\right) + \theta_2\left(\frac{V_{gsteff} + 2V_T}{T_{ox}}\right)^2. \qquad (13)$$

where $V_{gsteff}$ is the gate overdrive (which is ~ $V_G$-$V_T$ in strong inversion or when $V_G >> V_T$), $\theta_1$, $\theta_2$ are fitting parameters. In fact, $(V_{gsteff} + 2V_T)/T_{ox}$ in the right hand side of (13) effectively serves the purpose of $E_{eff}$ in the BSIM/SPICE model, where $V_{gsteff}$ and $2V_T$ terms reflect the effect of variation in $Q_{inv}$ and $Q_{dep}$, respectively. Moreover, the $\mu_0/(1+\alpha_{eff}N_{IT})$ factor in (12) reduces with increasing $N_{IT}$, to reflect the decrease in $\mu_{eff}@E_{eff}$ due to Coulomb scattering.

So according to (12)-(13), an increase in $N_{IT}$ not only reduces $\mu_{eff}$ through the $\mu_0/(1+\alpha_{eff}N_{IT})$ factor, but also reduces $\mu_{eff}$ through an increase in the $(V_{gsteff} + 2V_T)/T_{ox}$ factor of (13) . Therefore, $\mu_{eff}@V_G$ in the existing SPICE analysis always decreases with $N_{IT}$, irrespective of the $\mu_{eff}$-$E_{eff}$ steepness, which is inconsistent with our experimental observations in [19]. This discrepancy suggests that the variation of $E_{eff}$ due to $N_{IT}$ through the $(V_{gsteff} + 2V_T)/T_{ox}$ factor of (13) is completely unphysical in the existing SPICE model.[7] And as shown in the following paragraph, consideration of physical $\Delta E_{eff}$ has significant consequences regarding self-compensation, especially for strained technology with steep $\mu_{eff}$-$E_{eff}$ characteristics.

---

[7] In fact, $(V_{gsteff} + 2V_T)/T_{ox}$ term of (13) should fail to reflect the change in $E_{eff}$ due to almost all sources of threshold variation (except RDF and LER), where only $Q_{inv}$ or $V_{gsteff}$ changes with $\Delta V_T$.

*B.     Circuit Analysis using Physics-based $\Delta E_{eff}$*

Our experimentally calibrated physical $\mu_{eff}$ - $E_{eff}$ model, *i.e.* (4)-(7), enables us to simulate I-V characteristics before and after $N_{IT}$ generation, considering proper change in transistor parameters like $V_T$, $\mu_{eff}$, $E_{eff}$, *etc.* To perform these I-V simulations as a function of $V_G$ and $V_D$, we use the equations similar to the one used in PETE [48, 49]. The transistor parameters used in the I-V simulation are: $EOT$ = 1.2nm, $N_{Dop}$ = $10^{17}$ cm$^{-3}$, $T$ = 300 $^0$K, $L$ = 45nm, $W$ = 1μm, $/V_{T0}/$ = 0.4V (defined at $I_{D,lin} \sim 10^{-6}$A/μm), $I_{OFF}$ = 1nA/μm, $\lambda$ = 0.1, $SS$ = 106mV/dec, $DIBL$ = 107mV/V. In addition, we also use the classical SPICE-mobility model, *i.e.* (12)-(13), to perform similar I-V calculations using PETE. Fig. 7a,b clearly show the difference between the classical SPICE model (based on (12)-(13)) and the physical model (based on (4)-(7)) in predicting the change of $I_D$-$V_G$ (at $V_{DS}$ = -0.1V) due to $N_{IT}$-induced $\Delta V_T$. Mobility model in classical SPICE simulation [46] estimates an unphysical increase in $E_{eff}$ due to $N_{IT}$ generation and hence degrades $\mu_{eff}@V_G$ or $I_D@V_G$ significantly, especially for 2.5% strained transistor (Fig. 7b) that has more $\mu_{eff}$-$E_{eff}$ steepness. On the other hand, the physical mobility model considers the effect of partial self-compensation due to $N_{IT}$-induced $E_{eff}$ reduction and hence predicts significantly less I-V degradation for 2.5% strained transistor.

Next, we study the effect of $N_{IT}$-induced $\Delta V_T$ at $V_{DD}$ = 1V for (i) NAND gate driving an INV gate and (ii) 5-stage ring oscillator RO. These circuit configurations are further elaborated in [48]. Our calculation of delay degradation due to $N_{IT}$-induced $\Delta V_T$ (Fig. 7c,d) demonstrates the importance of using correct sign for $\Delta E_{eff}$ in the circuit analysis, especially for the strained technology (see Fig. 7d). While unphysical variation of $E_{eff}$ due to $N_{IT}$ in classical SPICE-mobility analysis predicts significant delay degradation with increase in strain, a physical mobility analysis predicts much less delay degradation for the strained technology. More importantly, although the $I_{D,lin}$-$V_G$ characteristics of 2.5% strained transistor has no flatness up to 1V (Fig. 7b), presence of partial self-compensation in strained transistor (through positive $\Delta\mu_{eff}$ in (1)) reduces delay degradation by ~15% at $\Delta V_T$ = 30mV (*e.g.*, changes NAND delay degradation from 4.8% for unstrained to 4.07% for strained transistor).

Thus our analysis with uniaxial strain silicon technology suggests the importance of physical mobility modeling in transistor/circuit level variability analysis. Although total self-compensation (zero delay degradation) is not shown for $V_{DD}$ = 1V at the practical limits of uniaxial strain (Fig. 7), the reduction of delay degradation with strain (albeit by only ~0.73%) suggests the importance of $\mu_{eff}$-$E_{eff}$ steepness in reducing the effect of CMOS variability. The recent reports of $I_D$-$V_G$ characteristics in III-V transistors [50] indicate the presence of large $\mu_{eff}$-$E_{eff}$ steepness. Therefore, these advanced CMOS technology may reduce the delay-degradation further, and hence result in considerable increase in IC lifetime [20, 21], only if $\Delta E_{eff}$ due to $\Delta V_T$ is correctly taken into account.

## VIII. CONCLUSION

We have demonstrated how the concept of variation resilience in existing MOS architecture is related to the negative steepness of the $\mu_{eff} - E_{eff}$ relationship, or equivalently to the presence of smaller $g_{m,ON}$. Such design is expected to make transistors less sensitive to various sources of $V_T$ variations, including those related to fluctuations in interfacial/oxide defects, oxide thickness, metal work-function, *etc.* Broadly speaking, the increase in $\mu_{eff} - E_{eff}$ steepness can compensate any $V_T$ fluctuations that arise due to parameters within the oxide and gate regions, as well as in the oxide/substrate interface. In addition, we have studied the extent of variation resilience in uniaxial strained CMOS technology and have shown partial presence of self-compensation in these transistors. This study has also identified the importance of physical mobility modeling in circuit analysis for advanced CMOS technology, by considering appropriate changes in transistor parameters (especially effective electric field) due to threshold voltage variation. Therefore, it is predicted that the effect of self-compensation will be observable in circuit/system level by further increase in the $\mu_{eff} - E_{eff}$ steepness (for example, using III-V transistor technology), provided that circuit analysis reflects physical changes in transistor parameters. This might substantially reduce the burden on guard-band voltage, expensive design algorithms and/or extra circuitry – that are currently being used for mitigating the effect from $V_T$ variations.

**Acknowledgement**

This work was done with financial support from TSMC, Nanoelectronic Research Initiative, 2008 IEEE EDS PhD Fellowship and 2009-2010 Intel Foundation PhD Fellowship. We gratefully acknowledge Dr. Khaled Ahmed (Applied Materials) for the strained transistors, Network for Computational Nanotechnology (NCN) for computational resources, and Birck Nanotechnology Center (BNC) for experimental facilities. In addition, we also acknowledge our discussions with Prof. Mark S. Lundstrom (Purdue University) and Prof. Kevin Cao (Arizona State University) on various related issues.

**Figure Captions:**

Fig. 1: (a) Linear transfer characteristics of a classical MOS transistor, where any sources of $\Delta V_T$ lead to $\Delta I_{D,lin}$. The effect of $\Delta V_T$ can be minimized by using guard-band voltage ($V_{DD}$-$V_I$). (b) Comparison of mobility-field ($\mu_{eff}$ – $E_{eff}$) relationship in classical and variation-resilient transistors. An increase in $V_T$ always results in an increase in $\mu_{eff}$, except for $N_{IT}$-related $\Delta V_T$. (c) Transfer characteristics of the proposed variation-resilient MOS transistor, where $\Delta V_T$ is not reflected in $\Delta I_{D,lin}$. Thus variation-resilient transistor can operate with reduced guard-band.

Fig. 2: Measured $\mu_{eff}$-$E_{eff}$ at different levels of strain ($\varepsilon$) are fitted using (4)-(7). Here, we estimate $\varepsilon$ by using Fig. 11 of [43]. The measurement at 0% strain is consistent with the universal mobility measurement (dashed line) of [22].

Fig. 3: (a) Calculation of $\Delta_{H,\varepsilon}$ within single sub-band, Airy-function approximation shows an increase in $\Delta_{H,\varepsilon}$ with $E_{eff}$, as also observed in [40]. (b) $\Delta E_{LH\text{-}HH} = \Delta_{H,\varepsilon} - \Delta_{H,\varepsilon=0}$ at different $E_{eff}$ is close to its magnitude of 6.426$\varepsilon$ eV at zero $E_{eff}$, where 6.426eV is the deformation potential.

Fig. 4: (a) $\mu_{eff}$-$E_{eff}$ curves for different levels of strain, estimated using strain-dependent mobility model of (4)-(7). (b) Corresponding $I_{D,lin}$-$V_G$ characteristics at $V_{DS}$ = 50mV for a simulated transistor having $L$ = 100μm, $EOT$ = 1.4nm, and $N_{Dop}$ = 3x10$^{17}$ cm$^{-3}$.

Fig. 5: (a) Comparison of $\mu_{eff}$ – $E_{eff}$ steepness for unstrained ($\varepsilon$ = 0%) and strained ($\varepsilon$ = 2.5%) transistors. (b, d) Unstrained transistor having less $\mu_{eff}$ – $E_{eff}$ steepness suffers from $EOT$ and $\Phi_{GS}$ variations. Whereas, (c, e) strained transistor with higher $\mu_{eff}$ – $E_{eff}$ steepness or small $g_{m,ON}$ is less sensitive to $EOT$ and $\Phi_{GS}$ variation at operating $V_G$.

Fig. 6: (a) Unstrained transistor suffers from ±20% $N_{Dop}$ variation. (b) Though $g_{m,ON}$ ~ 0 at $V_{GS}$ ~1.2-1.5V for 2.5% strained transistor, dopant fluctuation remains uncompensated around that range of $V_{GS}$.

Fig. 7: Simulated $I_D$-$V_G$ characteristics at $|V_{DS}|$ = 0.1V for (a) unstrained and (b) 2.5% uniaxial strained transistor. Effect of $N_{IT}$-induced $\Delta V_T$ in NAND gate and 5-stage RO circuits for (c) unstrained and (d) 2.5% uniaxial strained transistor. The difference between classical SPICE simulation based on (12)-(13) (solid symbols) and circuit simulation based on (4)-(7) (open symbols) is clearly evident in strained CMOS transistor.

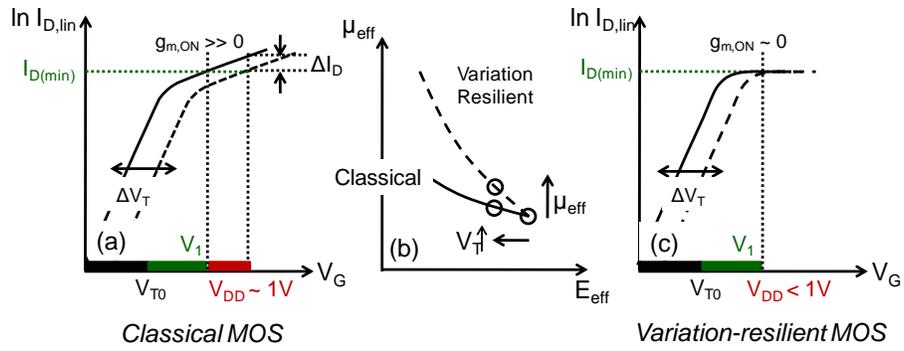

Fig. 1: Islam *et al.*

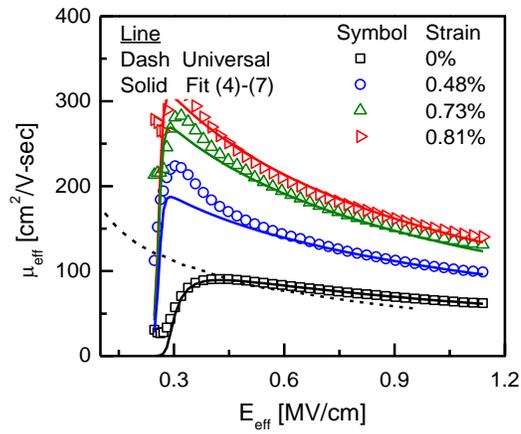

Fig. 2: Islam *et al.*

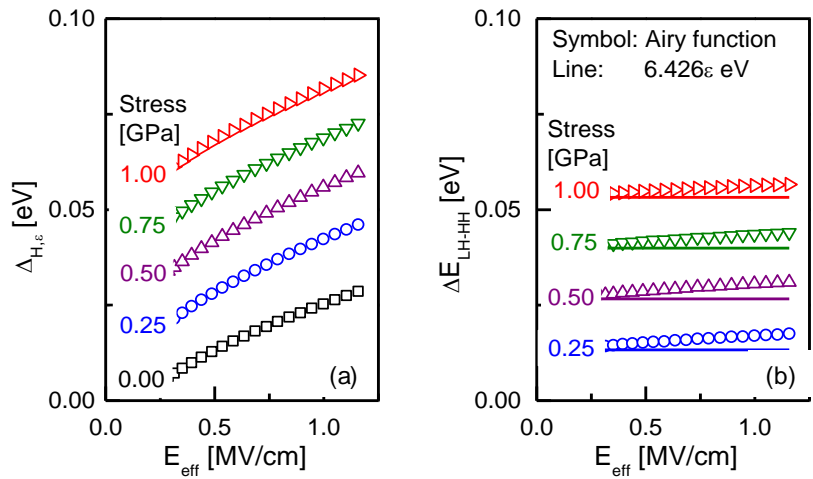

Fig. 3: Islam *et al.*

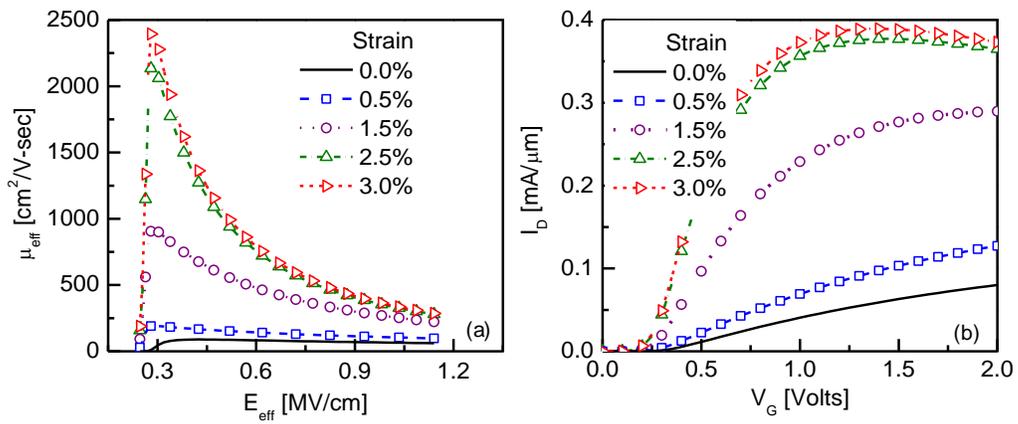

Fig. 4: Islam *et al.*

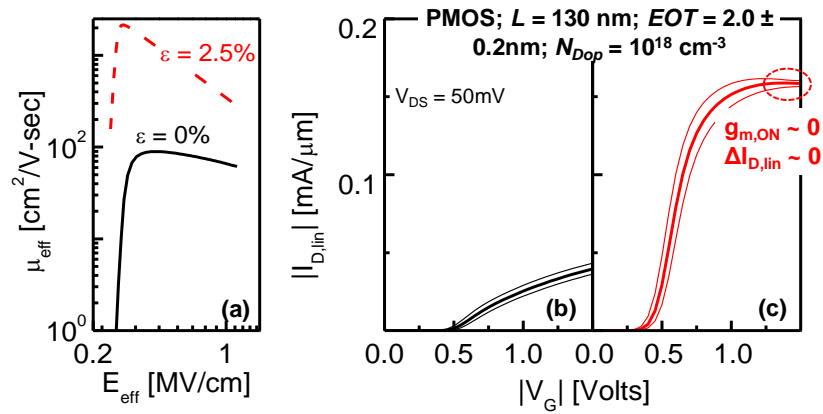

Fig. 5: Islam *et al.*

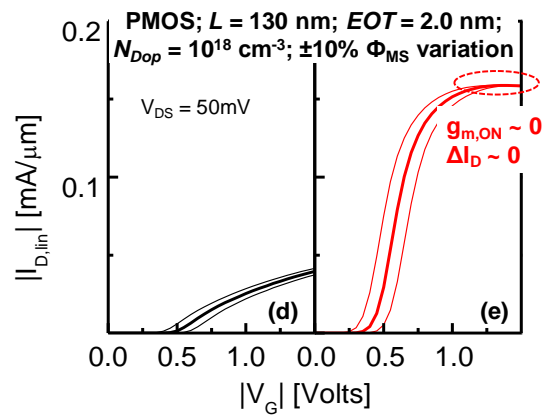

Fig. 6: Islam *et al.*

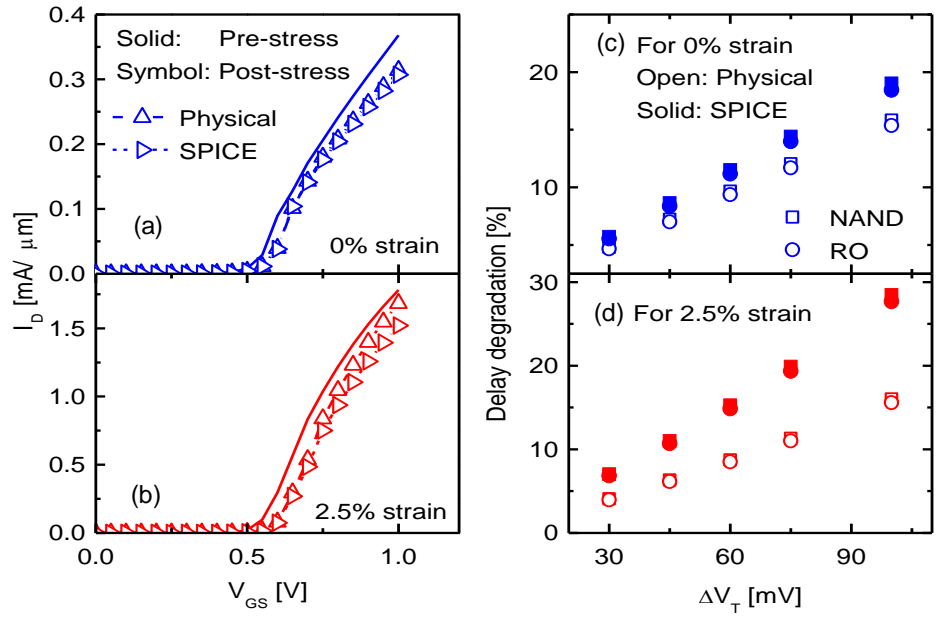

Fig. 7: Islam *et al.*